\documentclass[conference]{IEEEtran}
\IEEEoverridecommandlockouts
\usepackage{cite}
\usepackage{amsmath,amssymb,amsfonts}
\usepackage{algpseudocode}
\usepackage{algorithm}
\usepackage{graphicx}
\usepackage{booktabs}
\usepackage{subfig}
\usepackage{textcomp}
\usepackage{xcolor}
\usepackage{multirow}
\usepackage{makecell}
\usepackage{url}

\def\BibTeX{{\rm B\kern-.05em{\sc i\kern-.025em b}\kern-.08em
    T\kern-.1667em\lower.7ex\hbox{E}\kern-.125emX}}

\begin{document}

\title{A Robust Framework for Sybil Attack Detection in Vehicular Ad Hoc Networks
}

\author{\IEEEauthorblockN{Md. Sadmin Tahmid Khan, Md. Saim Ahmmed Utsho, Mosarrat Jahan}
\IEEEauthorblockA{Department of Computer Science and Engineering, University of Dhaka, Dhaka, Bangladesh\\
 Email: mdsadmintahmid-2019317818@cs.du.ac.bd, mdsaimahmmed-2019617824@cs.du.ac.bd, mosarratjahan@cse.du.ac.bd}}
\maketitle

\begin{abstract}
Sybil attacks create an illusion of traffic congestion by utilizing fake identities, which undermines the reliable and safe operation of vehicular ad hoc networks (VANETs). Existing detection mechanisms struggle to effectively handle Sybil attacks as they are (i) susceptible to high false positive rates (FPR) due to the overlapping trajectories of both Sybil and legitimate vehicles, (ii) not practical for real-world deployment due to manual calibrations with ground data, (iii) ineffective for sparse distribution of roadside units (RSUs) and vehicles as they depend heavily on the presence of both, and (iv) inefficient due to computational overheads. This paper addresses these shortcomings and proposes a robust framework to handle Sybil attacks. The proposed scheme reduces the FPR by utilizing Global Positioning System (GPS) location data, enabling the construction of more accurate and distinguishable trajectories. Besides, it employs DBSCAN clustering to identify Sybil vehicles, facilitating unsupervised parameter selection. GPS data eliminates the dependency on RSUs and vehicles, making this scheme effective in both sparse and dense regions. Additionally, the proposed scheme is lightweight and consistent across vehicles with heterogeneous capacities. Experimental results demonstrate that the proposed scheme reduces the FPR by approximately 72\% in both dense and sparse regions and lowers the false negative rate (FNR) by 52\% in sparse regions. Besides, it exceeds existing works by achieving a detection rate of 99\% in sparse regions. Additionally, the proposed scheme decreases the detection time by almost 76\% in dense regions and 43\% in sparse ones.
\end{abstract}


\section{Introduction}
The secure and systematic operation of VANETs plays a crucial role in preserving the integrity of Intelligent Transportation Systems (ITS). VANETs automate the traffic management system by enabling network connectivity among vehicles and Roadside Units (RSUs). Information exchanged among VANET entities is used to effectively manage the transportation system, leading to reduced traffic congestion, minimized road accidents, and improved road safety. Despite the significant potential of VANETs, they are highly susceptible to various security attacks due to their dynamic and time-sensitive nature. Such attacks ultimately compromise the reliability and effectiveness of VANETs, leading to chaotic road conditions, severe traffic congestion, and even life-threatening accidents. Hence, establishing robust security mechanisms for VANETs is essential, given their critical roles in transportation systems.

\begin{figure}
    \centering
    \includegraphics[width=.45\textwidth]{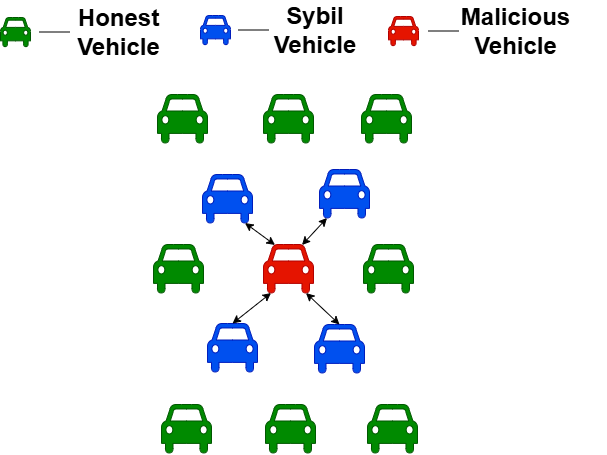}
    \caption{Sybil Attack in VANETs.}
    \label{fig_1}
\end{figure}

One such security attack that distorts the reliable operation of VANETs is the Sybil attack \cite{yu2013detecting}. In this attack, a malicious vehicle uses multiple identities to create the illusion of vehicles that do not exist in reality, known as a Sybil vehicle, as shown in Fig. \ref{fig_1}. A malicious vehicle may obtain multiple identities by using its pseudonyms (temporary, unlinkable identifiers to provide vehicle privacy) \cite{qu2015security}, illegally collecting pseudonyms from genuine vehicles \cite{zhang2023detection}, and by obtaining multiple On-board Units (OBUs) through collusion with the registration authority (RA) or by theft from other vehicles \cite{hasan2020securing, sakiz2017survey}.

A successful Sybil attack triggers Sybil nodes to broadcast false beacon messages, causing the illusion of traffic congestion to nearby vehicles \cite{zhang2015exploiting}, which manipulates the traffic management system to reroute them. This intentional rerouting causes real congestion on smaller roads, leading to delay, accidents, and obstructing emergency vehicles \cite{rabieh2015cross}. Besides, Sybil vehicles can interfere with the warning messages sent by legitimate vehicles, potentially endangering human lives \cite{pattanayak2020novel}. Moreover, Sybil attacks can be employed to gain unfair advantages in voting and reputation systems, significantly undermining security and trust within vehicular networks \cite{subramanian2020decentralized}. Hence, an effective mechanism to detect and mitigate the impact of Sybil attacks is essential.

Trajectory-based approaches to detect Sybil attacks assume that the trajectories of Sybil vehicles are likely to be similar to those of the malicious vehicles that originate them. Hence, comparing the trajectories identifies and eliminates Sybil attacks \cite{chang2011footprint, baza2020detecting}. However, this method also increases the possibility of a high false positive rate, as legitimate vehicles may follow the same route as Sybil vehicles \cite{chang2011footprint, baza2020detecting}. Existing trajectory-based schemes \cite{chang2011footprint, baza2020detecting, rajendra2024sybil} depend on heuristic parameters to distinguish Sybil trajectories from legitimate ones. However, the values of these heuristic parameters are environment-specific, requiring manual adjustments using labeled data prior deployment, making them impractical for real-world applications. Besides, some existing methods \cite{ rajendra2024sybil, dutta2013time}, use vehicle-to-vehicle (V2V) communication in construction of trajectories. V2V communication increases the likelihood of collusion attacks, where malicious vehicles cooperate to fabricate or manipulate location data, resulting in high false negative rates \cite{rajendra2024sybil}. Additionally, the high mobility of vehicles leads to short-lived V2V interactions, making both location verification and trajectory construction challenging in VANETs. Furthermore, schemes utilizing V2V interactions are only effective in dense traffic areas and struggle in sparse regions, where fewer vehicles are available to provide reliable location proofs \cite{rajendra2024sybil, dutta2013time}. Some recent works \cite{baza2020detecting, rajendra2024sybil,li2021trajectory} use computationally expensive puzzle-solving techniques to limit the number of generated trajectories.  These techniques provide unfair advantages to malicious vehicles equipped with high computational resources, leading to  inaccuracies in detection.

Our proposed scheme addresses the above-mentioned shortcomings that hamper the effectiveness of the existing Sybil attack detection mechanisms. The key contributions of this paper are as follows:

\begin{itemize}
\item A trajectory-based Sybil attack detection mechanism that uses uniquely identifiable GPS locations to construct vehicle trajectories.  This approach enhances accuracy and minimizes both false positives and false negatives by providing a clearer distinction between trajectories. Additionally, it prevents collusion attacks by eliminating  V2V communication and proves effective in both sparse and dense network environments.

\item The proposed scheme utilizes the DBSCAN clustering algorithm \cite{deng2020dbscan}, \cite{cavallaro2021analysis} to identify malicious and Sybil vehicles. This clustering method employs a fully unsupervised parameter selection mechanism, eliminating the need for manual heuristic calibration based on labeled ground-truth data. 

\item Experimental results demonstrate that the proposed scheme reduces the FPR by nearly 72\% in dense and sparse regions. Although the proposed scheme slightly lags in detection rate due to a higher FNR in the dense region, it outperforms existing works by reaching a detection rate of 99\% in sparse areas. In the sparse region, the proposed scheme reduces the FNR by almost 52\%. Additionally, the proposed scheme outperforms the existing works in terms of detection time, improving by approximately 76\% in dense areas and 43\% in sparse regions.
\end{itemize}

The remainder of the paper is organized as follows. Section \ref{related_work} reports the related literature on trajectory-based Sybil attack detection. Section \ref{preliminaries} describes the fundamental knowledge, and Section \ref{proposed_scheme} presents the system model and thoroughly describes the proposed scheme. Furthermore, Section \ref{security_analysis} discusses the proposed scheme's resiliency against different security attacks,  while Section \ref{experimental_results} highlights its effectiveness in detecting Sybil attacks. Finally, the paper concludes with some directions for future work in Section \ref{conclusion}.

\section{Related Works}
\label{related_work}
Chang et al. \cite{chang2011footprint} proposed a mechanism to detect Sybil attacks by utilizing vehicle trajectories. In this approach, each vehicle receives a message signed by the RSU it crosses. These consecutive signed messages are used to generate a trajectory for each vehicle. These trajectories are then filtered based on heuristic parameters: check window size and trajectory length limit. Consequently, a trajectory similarity graph is formed and the maximal clique within this graph is used to identify Sybil vehicles. Finding the optimal values for the heuristic parameters requires access to the labeled data before deployment, rendering the scheme unsuitable for real-world applications. Besides, this approach is susceptible to high FPR and FNR. Due to its dependency on RSUs for trajectory construction, it is only effective in dense urban regions; not practical in sparse regions with fewer RSUs. 

Baza et al. \cite{baza2020detecting} extended the work of Chang et al. \cite{chang2011footprint} by employing Proof of Work (PoW) and location information. In this scheme, each RSU generates a signed, time-stamped tag as evidence of the vehicle's position. Tags collected from multiple consecutive RSUs form the vehicle's trajectory. Additionally, each vehicle must solve a computationally expensive puzzle using the PoW algorithm. For a malicious vehicle to create multiple Sybil vehicles, it must successfully solve the puzzle for each trajectory at every RSU. This reduces the probability of creating multiple Sybil vehicles from a single vehicle. However, an attacker with higher computational power or with the aid of cloud services can solve multiple puzzles and create multiple trajectories, whereas a vehicle with resource constraints may not be able to generate its own trajectory. This discrepancy increases FPR and FNR. Similar to Chang et al's \cite{chang2011footprint} mechanism, this scheme is also not suitable for a sparse vehicular environment with few RSUs. This scheme also requires access to labeled data before deployment to determine the optimal values for heuristic parameters.

\begin{table*}
\centering
\caption{Comparison of the proposed scheme with existing works}
\scalebox{0.75}{
\begin{tabular}{ |p{2.5cm}|p{3cm}|p{3cm}|p{2.9cm}|p{2.2cm}|p{3 cm}| }
 \hline
 \textbf{Scheme} & \textbf{Trajectory Construction} & \textbf{Detection Technique} & \textbf{Computationally Expensive?} & \textbf{Environment-specific Calibration?} & \textbf{Effective in Sparse / Dense regions?}\\
 \hline
  \hline
Chang et al. \cite{chang2011footprint} & V2R & Max clique & No puzzle solving & Yes & Dense deployment of RSUs \\
 \hline
Baza et al. \cite{baza2020detecting} & V2R & PoW and max clique & PoW puzzle solving & Yes & Dense deployment of RSUs\\ 
\hline
Li et al. \cite{li2021trajectory} & V2R & Puzzle chain and threshold signature & Chain of computational puzzles solving & Yes & Dense deployment of RSUs\\
\hline
Rajendra et al. \cite{rajendra2024sybil} & V2R and V2V & VDF chain and max clique & Use VDF & Yes & Dense deployment of RSUs and vehicles \\  
\hline
Dutta and Chellappan \cite{dutta2013time} & V2V & Fuzzy time-series clustering & No puzzle solving & No & Dense deployment of vehicles \\
\hline
Proposed Scheme & V2R (verification only) and GPS & DBSCAN clustering &  No Puzzle solving & No & Both sparse and dense deployment of RSUs and vehicles \\
\hline
\end{tabular}}
\label{tab_2}
\end{table*}

Li et al. \cite{li2021trajectory} addressed the issue of the unfair advantage obtained from using cloud resources to solve puzzles in Baza et al.’s \cite{baza2020detecting} scheme. In this method, similar to Baza et al.’s approach \cite{baza2020detecting}, a vehicle constructs its trajectory using signed, time-stamped tags issued by RSUs. Between RSUs, the vehicle solves a chain of computational puzzles, with each solution signed by the vehicle's secret key to prevent powerful cloud resources from solving multiple complex problems. An Announcement Manager checks both the correctness of the puzzle solutions and the validity of the signed RSU tags to verify the trajectory construction. Additionally, it evaluates the trajectory for unrealistic patterns and detects Sybil vehicles. The scheme employs only a single layer of Sybil attack detection, whereas Baza et al.’s scheme \cite{baza2020detecting} uses two layers: heuristic-based checking and a maximal clique method. Fewer detection layers lead to higher FPR and lower detection accuracy.

Rajendra et al. \cite{rajendra2024sybil} addressed the shortcomings of Baza et al.’s \cite{baza2020detecting} scheme by incorporating Verifiable Delay Functions (VDFs) with both V2V and Vehicle-to-RSU (V2R) interactions. VDFs ensure more consistent computation times across devices with varying processing capabilities, thereby reducing the advantage of malicious vehicles equipped with powerful hardware for solving difficult puzzles and generating multiple trajectories. Furthermore, the inclusion of V2V interactions helps create more distinctive trajectories, improving detection accuracy in areas with limited RSU coverage. However, V2V communication also introduces the risk of collusion among malicious vehicles. Moreover, the effectiveness of the scheme decreases in sparse traffic environments with fewer vehicles, leading to higher FPR and FNR. Similar to Chang et al.’s \cite{chang2011footprint} scheme, this approach relies on labeled data to initialize heuristic values.

Dutta and Chellappan \cite{dutta2013time} proposed a fuzzy time-series-based clustering mechanism to detect Sybil attacks. In this scheme, trajectories are constructed based on peer reporting, utilizing V2V communication. Each trajectory is created using the periodic location updates from other vehicles. These trajectories are grouped using the Fuzzy Short Time-Series (FSTS) clustering algorithm. A theoretical probability is used to determine whether the nodes are legitimate or Sybil. One major drawback of this scheme is the use of V2V to create trajectories. Vehicles may collude to report false locations, leading to higher FNR. This method is also inefficient in sparse regions with few vehicles.

In summary, existing trajectory-based Sybil attack detection mechanisms are susceptible to high FPR and FNR. Moreover, the inclusion of V2V communication in trajectory construction introduces several challenges. It increases the possibility of vehicle collusion, which can further increase the false negative rate. Additionally, the dependence on V2V communication makes existing solutions less effective in sparse traffic environments. Several schemes also employ computationally intensive mechanisms to restrict the generation of Sybil vehicles by malicious entities, thereby imposing additional computational overhead on vehicles. 

In contrast, the proposed scheme utilizes GPS location data to construct vehicle trajectories, which improves detection accuracy while reducing FPR and FNR. Furthermore, the proposed scheme avoids the use of V2V communication during trajectory construction, thereby eliminating the risk of collusion attacks. Despite not relying on V2V communication, the scheme remains effective in sparse regions due to its use of GPS data. Additionally, the proposed scheme does not employ computationally expensive mechanisms to limit the generation of Sybil vehicles, making it lightweight and suitable for vehicles with varying processing capabilities. It is also more practical for real-world deployment, as it eliminates the need for manual calibration during the detection process. Table \ref{tab_2} presents a comparison of the proposed scheme with existing approaches in the literature.

\section{Preliminaries}\label{preliminaries}
\subsection{Fr\'echet Distance}
The Fr\'echet distance is a spatio-temporal similarity measure used to compare two trajectories that takes into account the location and time ordering \cite{su2020survey}. 

Given two polygonal curves $P$ and $Q$ where $P = (u_1, u_2, u_3, \ldots, u_p)$ and $Q = (v_1, v_2, \ldots, v_q)$, and $d(u_i, v_j)$ is the Euclidean distance between two points of the curves, the discrete Fr\'echet distance can be defined as a dynamic programming problem. The discrete Fréchet distance between $P$ and $Q$ is then defined following \cite{EiterM94} as follows:
\begin{center}
$\delta_{dF}(P, Q) = ca(p, q)$
\end{center}
where $ca(i,j)$ =

\vspace{0.5em}
\scalebox{.90}{
$\begin{cases}
d(u_1, v_1), & \text{if } i = 1 \text{ and } j = 1, \\
max(ca(i-1, 1),\, d(u_i, v_1)), & \text{if } i > 1 \text{ and } j = 1, \\
\max(ca(1, j-1), \, d(u_1, v_j)), & \text{if } i = 1 \text{ and } j > 1, \\
\max( \min\{ ca(i-1,j), \, ca(i-1,j-1), \\ca(i,j-1) \}, \, d(u_i, v_j)), & \text{if } i > 1 \text{ and } j > 1.
\end{cases}$}

\subsection{DBSCAN Clustering Algorithm}\label{dbscan_clustering}
Density-based spatial clustering of applications with noise (DBSCAN) is an unsupervised clustering algorithm that groups data points based on density connectivity \cite{deng2020dbscan}. It does not require prior knowledge of the number of clusters and can effectively identify noise (outliers).

This algorithm works based on two key parameters:
\begin{itemize}
    \item $\varepsilon$ (epsilon): The maximum distance between two points to be considered as neighbors.
    \item $MinPts$: The minimum number of points within an $\varepsilon$-neighborhood to form a dense region (a cluster).
\end{itemize}

DBSCAN follows the density reachability principle \cite{ester1996density}, where a point $q$ is directly density-reachable from $p$ if:

\begin{equation}
\text{dist}(p, q) \leq \varepsilon \quad \text{and} \quad |N_\varepsilon(p)| \geq \textit{MinPts}
\end{equation}

where $N_\varepsilon(p)$ is the set of points within the $\varepsilon$-radius of $p$.

 The DBSCAN algorithm \cite{cavallaro2021analysis} starts with an unvisited point $p$, marks it visited, and finds its $\varepsilon$-neighborhood. If the number of neighbors $< MinPts$, $p$ is provisionally classified as an outlier. Otherwise, it is a core point, and a new cluster is initiated by including the core point $p$. All unvisited neighbors of the core point $p$ are visited, and their $\varepsilon$-neighborhoods are determined. If a neighbor $q$ of $p$ is a core point, then $q$ along with its neighborhood is also included in the cluster. If $q$ has less then $MinPts$ neighbors and is not part of any other clusters, only $q$ is added to the cluster. The cluster expansion continues until no additional points can be included. After a cluster is fully determined, the process continues with the next unvisited point. The DBSCAN algorithm terminates when all points are visited. After processing each point, any point that does not belong to a cluster is identified as an outlier.

 \section{Proposed Scheme}\label{proposed_scheme}
\subsection{System Model}
\begin{figure}
    \centering
    \includegraphics[width=.45\textwidth]{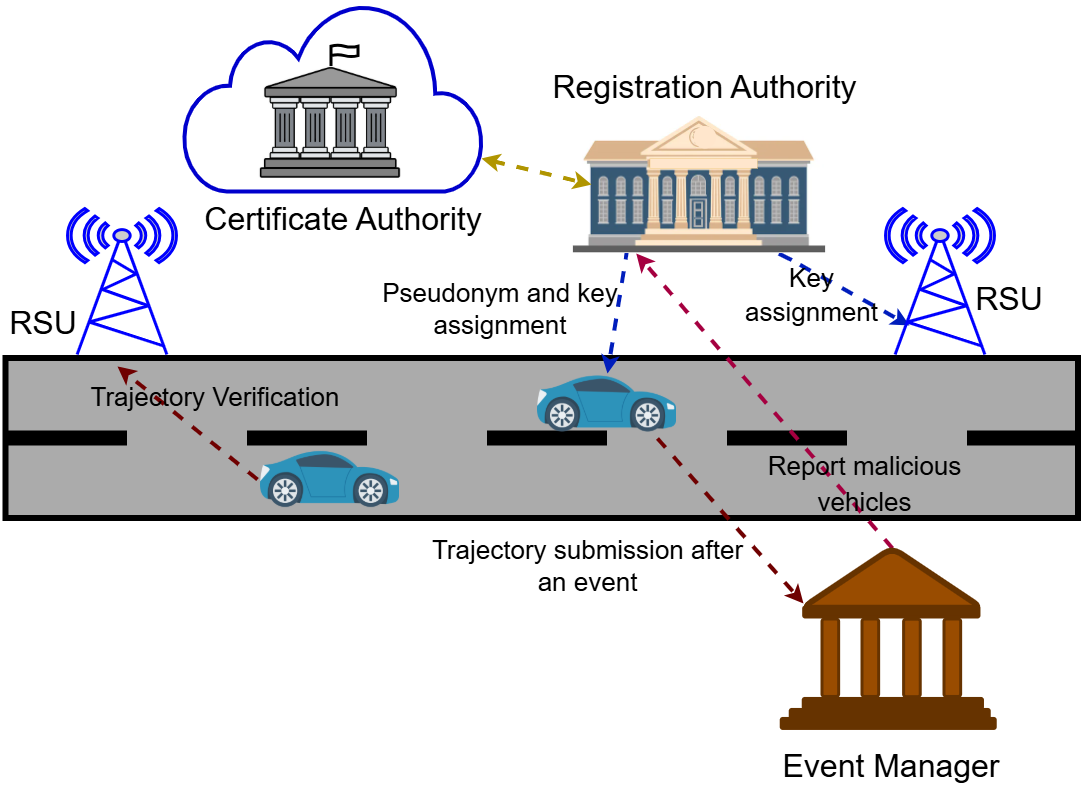}
    \caption{System model.}
    \label{fig_2}
\end{figure}

As shown in Fig. \ref{fig_2}, the system model comprises a \textit{Registration Authority (RA)}, \textit{Roadside Units (RSUs)}, \textit{vehicles} and \textit{Event Manager (EM)}. 

The \textit{registration authority} is a trusted entity under the Government's control, responsible for registering vehicles and issuing them pseudonyms. Additionally, it assigns and manages cryptographic keys to VANET entities such as \textit{RSUs} and \textit{vehicles} in collaboration with a Certificate Authority (CA) to ensure secure communication. It also identifies and removes malicious \textit{vehicles} from the network after receiving information from the \textit{event manager} and \textit{RSUs}.  \textit{Roadside units} are trusted edge devices usually placed at road junctions \cite{kara2020effect}. They provide storage, computation, and communication facilities to \textit{vehicles} \cite{wei2019rsu}, \cite{kperiyarselvam2024}, \cite{mershad2012roamer}. An \textit{RSU} receives a public and private key pair from a \textit{CA} through the \textit{RA}. They are responsible for verifying vehicle trajectories and reporting \textit{vehicles} with inappropriate trajectory construction to the $RA$. \textit{Vehicles} are moving entities equipped with on-board units (OBUs), GPS navigation systems, radars, accelerometers, cameras \cite{kaushik2022transmit}, communication modules, and GPS receivers \cite{zhao2016high}. The GPS receiver enables \textit{vehicles} to determine their precise location. Each \textit{vehicle} receives a public and private key pair from the \textit{CA} via \textit{RA}. \textit{Vehicles} are untrusted entities and may launch Sybil attacks.  
An \textit{event manager} is a trusted entity that actively monitors and analyzes the vehicle behavior to identify potential malicious behavior. It collects trajectory information from \textit{vehicles} after an event (a defined period of time when a \textit{vehicle} constructs a trajectory) and analyzes this data for patterns to identify malicious and Sybil \textit{vehicles}. Once a malicious or Sybil \textit{vehicle} is detected, the \textit{EM} reports the findings to the \textit{RA} for taking appropriate actions.  

\begin{figure}
    \centering
    \includegraphics[width=.3\textwidth]{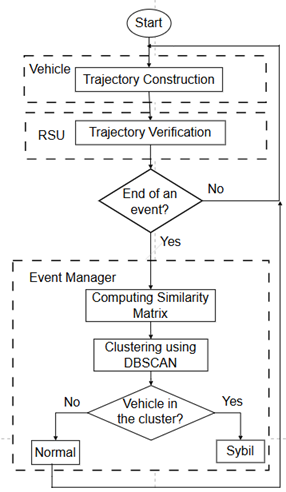}
    \caption{Workflow of the proposed scheme.}
    \label{fig_3}
\end{figure}

\subsection{Overview of the Proposed Scheme}
As shown in Fig. \ref{fig_3}, each vehicle constructs a trajectory representing its travel path over a fixed duration of time, known as an event. The trajectory consists of a sequence of GPS coordinates (latitude and longitude) and timestamps obtained from GPS signals via GPS receivers. When a vehicle passes by an RSU, it submits its trajectory to the RSU, which verifies and signs these trajectories to ensure accurate construction of paths. After the completion of an event, vehicles submit their signed trajectories to the event manager. The EM then preprocesses the data and creates a time series for each vehicle, representing the latitude and longitude coordinates over time. It measures the similarity between pairs of trajectories using the discrete Fr\'echet distance metric and constructs a similarity matrix. Finally, the EM applies DBSCAN clustering \cite{deng2020dbscan}, \cite{cavallaro2021analysis} to the data to form clusters, which identifies patterns in vehicle movement and detects Sybil vehicles. The notations used to describe the proposed scheme are shown in Table \ref{table_1}.
\begin{table} 
\centering
\caption{List of Notations}
\begin{tabular}{|l|p{6.5cm}|}
\hline
\textbf{Notation}                 & \textbf{Description}     \\ \hline
$T_{ij}$ & Trajectory of a vehicle from $RSU_i$ to $RSU_j$\\ \hline
$S_i$ & A seed value generated by $RSU_i$\\ \hline
$t_i$   & Timestamp of vehicle passing by $RSU_i$\\ \hline
$I_i$   & Unique identifier for $RSU_i$\\ \hline
$tag_i$   & Concatenation of $t_i$ and $I_i$\\ \hline
$sk_i$   & Secret key of $i$th RSU or vehicle\\ \hline
$\sigma_{i}$  &  Signature of $i$th RSU or vehicle\\ \hline   
$Cert_i$ & Certificate of $i$the RSU or vehicle\\ \hline
$pk_i$  & Public key of $i$th RSU or vehicle\\ \hline   
\end{tabular}
\label{table_1}
\end{table}

\subsection{Trajectory Construction}
\begin{figure}
    \centering
    \includegraphics[width=.51\textwidth]{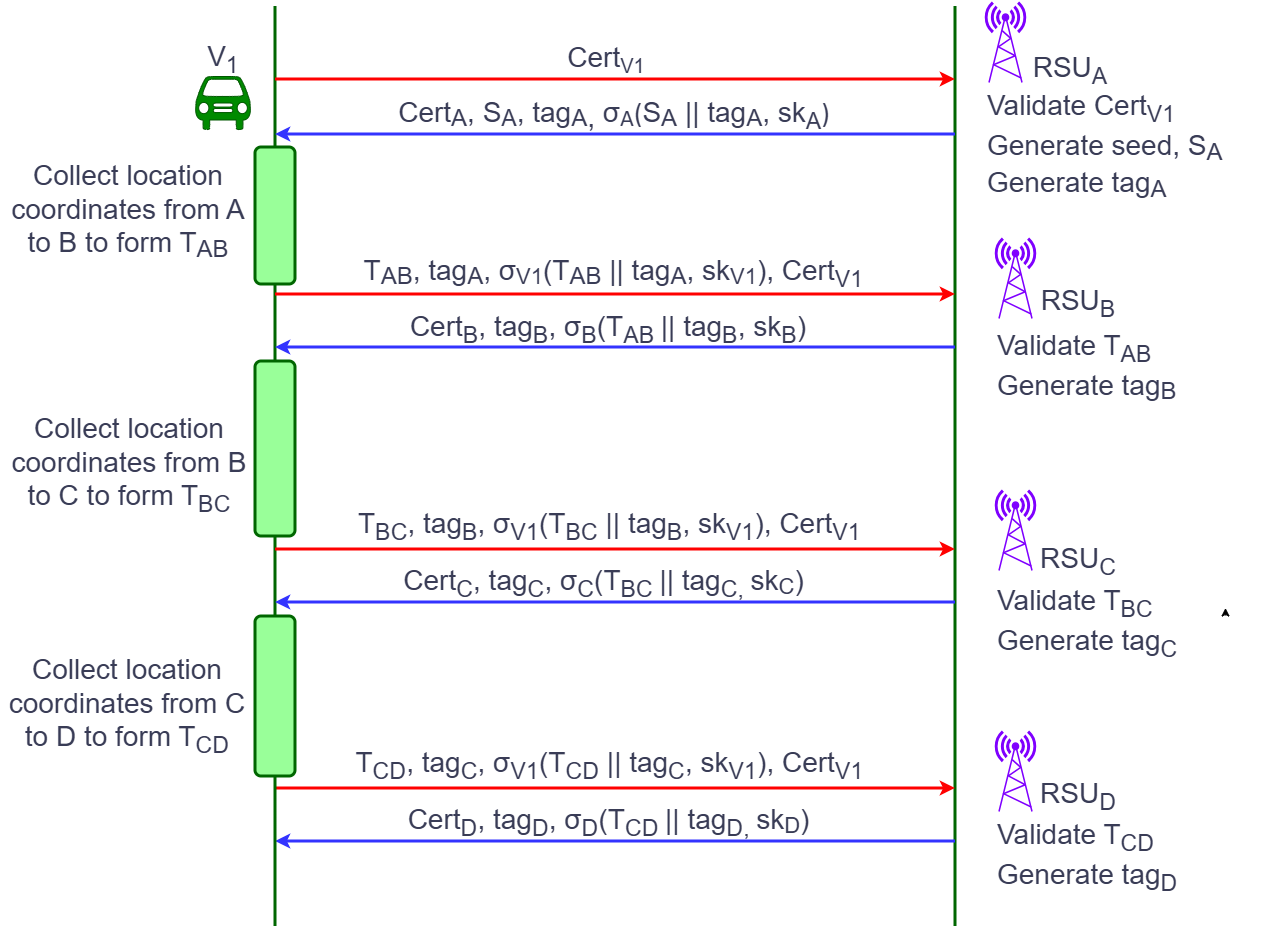}
    \caption{$V_{1}$'s interaction and message exchanges with $RSU_{A}$, $RSU_{B}$, $RSU_{C}$ and $RSU_{D}$. }
    \label{fig_4}
\end{figure}

To illustrate the operation of this phase, an example shown in Fig. \ref{fig_4} is employed. In this example, a vehicle $V_1$ constructs its trajectory while traveling through $RSU_{A}$, $RSU_{B}$, $RSU_{C}$, and $RSU_{D}$.

\begin{itemize}
\item $V_{1}$ starts its journey from $RSU_{A}$ by sending its certificate, $Cert_{V_1}$ for verification. Once $V_1$ is authenticated, $RSU_{A}$ generates a seed, $S_{A}$ and a tag, $tag_{A}$ containing a timestamp $t_{A}$ of the interaction and a unique identifier $I_{A}$ of ${RSU_A}$.
\[ tag_A : t_A || I_A \]
$RSU_{A}$ then signs both $tag_{A}$ and $S_{A}$ using its private key, $sk_{A}$.
\[\sigma_A = \sigma(S_{A} || \text{tag}_A, sk_A)\]
Finally, $RSU_{A}$ sends $Cert_A$, $tag_{A}$, $S_{A}$, and $\sigma_{A}$ to $V_{1}$. 
\item As $V_{1}$ continues its journey, it forms a trajectory $T_{AB}$ (when a vehicle is visiting from $RSU_A$ and $RSU_B$). Each row of $T_{AB}$ corresponds to a vector of the \textit{timestamp, latitude}, and \textit{longitude}, representing the data recorded at a particular timestamp throughout the vehicle's journey. 

When $V_{1}$ begins its journey from $RSU_{A}$ towards $RSU_B$, it starts forming $T_{AB}$ by collecting data at different time point. Each data point is recorded whenever the vehicle receives a beacon message from the nearest RSU. Vehicle $V_1$ then signs the trajectory $T_{AB}$ along with $tag_A$ using its private key $sk_{V_1}$. It then sends $T_{AB}$, $tag_A$, $\sigma_{V_1} (T_{AB}||tag_A, sk_{V_1})$, and $Cert_{V_1}$ to $RSU_{B}$ for verification.

\item $RSU_{B}$ verifies $Cert_{V_1}$ and obtains $V_1$'s public key $pk_{V_1}$. It then verifies the submitted trajectory $T_{AB}$ and $tag_{A}$ using $\sigma_{V_1}$ and $pk_{V_1}$. Once the verification is successful, it then proceeds to perform location verification as discussed in Section \ref{trajectory_verification}.

\item If $RSU_B$ successfully verifies the locations in  $T_{AB}$, it signs both $T_{AB}$ and $tag_B$ with its private key $sk_B$.
\vspace{0.5em}

\hspace{5 em} $\sigma_B = \sigma (T_{AB} || tag_B, sk_B)$

\vspace{0.5em}
where $tag_B$ = $t_B || I_B$, $t_B$ is the timestamp when $V_1$ encounters $RSU_B$ and $I_B$ is the identifier of $RSU_B$.

Finally, $RSU_B$ sends $Cert_B$, $tag_B$  and $\sigma_B$ to the vehicle $V_1$.

\item $V_1$ validates the received $Cert_B$ and obtains the public key $pk_B$ of $RSU_B$. It then verifies $\sigma_B$ using the received $tag_B$ and $T_{AB}$ kept in its custody. If the $\sigma_B$ is valid, $V_1$ proceeds to construct the next trajectory $T_{BC}$.

\item As $V_1$ continues its journey, it progressively constructs its trajectory between each pair of RSUs. $V_1$ signs the trajectory using its private key and then sends the signed trajectory to the next RSU.

\end{itemize}

When an event expires, vehicles submit their trajectories to the Event Manager for further analysis. In this  example, the event concludes after \( V_1 \) encounters \( RSU_D \).

\subsection{Trajectory Verification}\label{trajectory_verification}
\begin{figure}
    \centering
    \includegraphics[width=.5\textwidth]{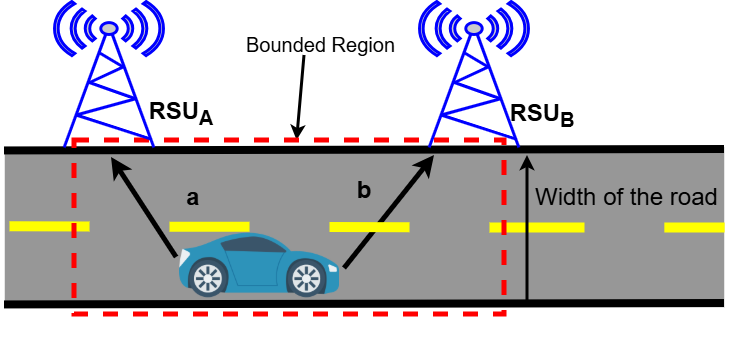}
    \caption{Area bounded for location verification.}
    \label{fig_5}
\end{figure}

When a vehicle submits $T_{AB}$ and $tag_A$, $RSU_B$ retrieves the previous RSU interaction using $tag_A$ and locates the position of the previous RSU by querying RA with the identifier $I_A$. It then verifies the submitted trajectories using Algorithm \ref{algo_1} that checks whether each $(x, y) \in T_{AB}$ lies with in a bounded region (lines 10 $\sim$ 14) using Algorithm \ref{algo_2}. The bounded region is defined as a rectangular area between the current RSU’s location $(X_B, Y_B)$ and the previous RSU’s location $(X_A, Y_A)$, with the width of the road as shown in Fig \ref{fig_5}. The x-coordinate of the vehicle is checked to see if it lies between $X_A$ and $X_B$ and the y-coordinate of the vehicle is checked to see if it lies within the road (line 5 of Algorithm \ref{algo_2}). In this algorithm, we assume that the both RSUs are placed in one side of the road. If fewer than 90\% of the trajectory points fall within the bounded region (line 15 of Algorithm \ref{algo_1}), the $RSU_B$ determines that the location data has likely been tampered with, indicating potential falsification of trajectory information. The 90\% threshold accounts for typical GPS inaccuracies, sensor errors, and environmental factors that can cause minor deviations in vehicle location data \cite{liu2024joint}.

\begin{algorithm}
\caption{Algorithm to verify trajectory at $RSU_B$}
\label{algo_1}
\begin{algorithmic}[1]
\Require
\Statex $T_{AB}$ $\gets$ Trajectory data between $RSU_A$ and $RSU_B$
\Statex $tag_{A}$ $\gets$ Tag of previous RSU
\Statex $width$ $\gets$ width of the road segment
\Ensure 
\Statex $Flag$ $\gets$ True (if location is valid)/False (if location is spoofed)
\Statex
\State $List\_X[] \gets get\_lattitudes(T_{AB})$
\State $List\_Y[] \gets get\_longitudes(T_{AB})$
\State $X_B \gets get\_lattitude(RSU_B)$
\State $Y_B \gets get\_longitude(RSU_B)$
\State $R_A \gets get\_RSU(tag_A)$
\State $X_A \gets get\_lattitude(R_A)$
\State $Y_A \gets get\_longitude(R_A)$
\Statex
\State $count \gets 0$
\State $Flag \gets true$ 

\For{$(x, y)$ in $(List\_X, List\_Y)$}
    \If{$bounded\_reg(x, y, X_A, Y_A, X_B, Y_B, width)$}
        \State $count \gets count + 1$
    \EndIf
\EndFor
\If{${count}/{len(T_{AB})} < 90\%$}
    \State $Flag$ $\gets$ false 
\EndIf
\State \Return $Flag$
\end{algorithmic}
\end{algorithm}

\begin{algorithm}
\caption{Algorithm to check coordinates in bounded region}
\label{algo_2}
\begin{algorithmic}[1]
\Require
\Statex $(x, y)$ $\gets$ coordinates in $T_{AB}$
\Statex $(X_A, Y_A)$: coordinates of $RSU_A$
\Statex $(X_B, Y_B)$: coordinates of $RSU_B$
\Statex $width$: width of the road
\Ensure
\Statex $Flag$ $\gets$ True (if within bounded region)/False otherwise.
\Statex
\State $minX \gets min(X_A, X_B)$
\State $maxX \gets max(X_A, X_B)$
\State $minY \gets min(Y_A, Y_B)$
\State $maxY \gets minY + width$

\If{$(minX \leq x \leq maxX) \And (minY \leq y \leq maxY)$}
    \State $Flag$ $\gets$ True 
    \Else
    \State $Flag$ $\gets$ false
\EndIf
\State \Return $Flag$
\end{algorithmic}
\end{algorithm}

To further verify the authenticity of the submitted trajectory, $RSU_B$ calculates the distances between each trajectory point and the two RSUs, denoted as $RSU_A$ (previous RSU) and $RSU_B$ (current RSU). For each point in the trajectory $T_{AB}$, the distance $b$ represents the distance of $RSU_B$ from the vehicle’s position, while $a$ represents the distance from $RSU_A$ to the vehicle’s position as shown in \ref{fig_5}. These distances are computed for all points in $T_{AB}$, and their order is analyzed. The values of $b$ should exhibit a decreasing order as the vehicle approaches $RSU_B$, whereas the values of $a$ should follow an increasing order as the vehicle moves away from $RSU_A$. This additional check ensures that the reported movement follows a logical and continuous trajectory, preventing location spoofing.

Vehicles that fail location verification are considered malicious. These vehicles are reported to the $RA$ and they are not further considered for trajectory analysis by the $EM$.

\subsection{Trajectory Analysis}
At the end of an event, $V_1$ submits its collected data to the $EM$. For Fig. \ref{fig_4}, data includes tags $\{tag_A, tag_B, tag_C, tag_D\}$, individual trajectories $\{ T_{AB}, T_{BC}, T_{CD} \}$, the initial seed $S_A$, and corresponding signatures $\{ \sigma_A, \sigma_B, \sigma_C, \sigma_D \}$. The $EM$ uses tags to identify RSUs and uses their public keys (obtained through certificates) to verify the signatures.

\subsubsection{Measurement of Similarity}
The $EM$ combines the trajectory segments received from a vehicle according to the order of the RSU placement to form the vehicle’s complete trajectory. The combined trajectory represents the GPS data that has already been collected for different timestamps and verified by RSUs. After collecting trajectories from all vehicles, the similarity value is computed between each pair of vehicle trajectories using the discrete Fr\'echet distance \cite{EiterM94}, capturing the temporal aspect by adding a time dimension to the spatial dimensions \cite{gutschlag2022generalized}. Besides, trajectory data varies in length due to differences in sampling rates and the duration of vehicle movements. The discrete Fr\'echet distance between two trajectories can be computed even if they have different numbers of points. Pre-processing trajectories to a fixed length is not required.

\subsubsection{Similarity Matrix Construction}
The $EM$ constructs a $N \times N$ similarity matrix $S$, where each entry $S [i][j]$ represents the computed discrete Fr\'echet value between trajectories $i$ and $j$, $1 \leq i,j \leq N$, and $N$ is the number of vehicles. 

\subsubsection{Cluster Construction} The $EM$ forms clusters using the DBSCAN clustering algorithm \cite{cavallaro2021analysis} to identify Sybil vehicles. This clustering method is used to identify dense clusters of Sybil vehicles using the similarity matrix $S$. To utilize the DBSCAN algorithm, the proposed scheme sets \textit{MinPts} to $2$, considering that a malicious vehicle generates at least one Sybil vehicle, forming a cluster of two points. To determine the optimal $\varepsilon$ value for DBSCAN, we use the elbow method \cite{schubert2017dbscan}. This approach computes the average distance of each point to its $k-1$ nearest neighbors and sorts these distances in ascending order. The sorted values are then plotted on a graph, where the optimal $\varepsilon$ is identified at the point of maximum curvature, known as the elbow point \cite{schubert2017dbscan}, representing a sharp change in the rate of distance increase. For $\varepsilon$-neighborhood calculation, the proposed scheme considers the discrete Fr\'echet distance between trajectories belongs to $S$. The parameter $k$ is generally set equal to the \textit{MinPts} value \cite{rahmah2016determination}; thus, for our scheme $k = 2$.

Once the optimal $\varepsilon$ is determined, the proposed scheme executes the DBSCAN clustering algorithm \cite{cavallaro2021analysis} on the collection of vehicle trajectories. Vehicles that are part of clusters are considered Sybil and outliers are classified as legitimate vehicles. Vehicles classified as Sybil vehicles include both the original malicious vehicle and the Sybil identities it generated. The $EM$ reports these identified vehicles to the $RA$ for further investigation and necessary action.

\section{Security Analysis}\label{security_analysis}
The proposed scheme is resistant to the attacks discussed in this section.

\subsection{Location Spoofing}
Trajectory-based Sybil attack detection mechanisms require constructing trajectories using a sequence of vehicles' location data. The basis of the trajectory-based detection mechanism is that Sybil vehicles follow similar trajectories. By providing false location data through location spoofing, vehicles can disrupt the functions of the Sybil attack detection mechanism. The proposed scheme prevents location spoofing using a strict verification process. Firstly, each RSU checks the GPS data to determine whether the points lie within the bounded region. Moreover, the data is checked for movement consistency. If any of these checks fail, the RSU flags the vehicle as potentially malicious and reports it to the $RA$ for further investigation, effectively preventing location spoofing.

\subsection{Collusion Attack}
A collusion attack occurs when vehicles cooperate to launch an attack on the network. In some trajectory-based schemes, V2V communication is employed, where one vehicle signs the location of another as part of trajectory formation \cite{rajendra2024sybil}. However, this step opens up the possibility of collusion, where vehicles can falsify location data by validating each other's false positions. In contrast, the proposed scheme deploys RSUs, a trusted entity, to verify and sign the trajectories generated by individual vehicles. Hence, the proposed scheme effectively resists vehicle collusion by avoiding V2V communication.

\subsection{Masquerading Attack}
In earlier works, a malicious vehicle can impersonate a legitimate vehicle by using the legitimate vehicle’s trajectory and claiming it as its own \cite{chang2011footprint}, \cite{baza2020detecting}. This impersonation enables malicious vehicles to mislead the Sybil attack detection mechanism, resulting in an increased false negative rate. The proposed scheme addresses this problem by utilizing a public and private key pair for each vehicle. Each vehicle signs its trajectory with its private keys, which can be verified by RSUs using the associated public key. Thus, each trajectory is uniquely associated with the vehicle that generated it. Additionally, the proposed scheme utilizes GPS location information to generate trajectories, making it challenging for a malicious vehicle to predict and replicate these patterns, which increases the difficulty of executing masquerading attacks.

\subsection{Pseudonym Abuse}
In VANETs, each vehicle is assigned a set of pseudonyms by $RA$ to hide its real identity to ensure privacy. Malicious vehicles can misuse these pseudonyms to create fake Sybil vehicles, leading to pseudonym abuse. The proposed scheme detects both malicious and Sybil vehicles. Once identified, these vehicles are reported to the $RA$ for appropriate action. Thus, the proposed scheme effectively addresses the issue of pseudonym abuse.

\subsection{Identity Theft}

Identity theft is a mechanism to launch Sybil attacks, where a malicious vehicle uses the identities of other vehicles to create Sybil vehicles. This may occur by stealing information from other vehicles and through the use of multiple OBUs in a single vehicle, obtained through unauthorized means, such as theft from other vehicles \cite{hasan2020securing}. The proposed scheme addresses this problem by detecting malicious and Sybil vehicles. Once these malicious vehicles are identified, they are reported to the $RA$ to handle the issues of identity theft.

\section{Experimental Evaluation}\label{experimental_results}

\subsection{Simulation Setup}
To evaluate the performance of the proposed scheme, we simulated a vehicular network. We extracted data from an area of $35$ $km \times 25$ $km$ from the map of Dhaka, the capital city of Bangladesh, and its surrounding area using OpenStreetMap \cite{map2017open}. This extracted map was then used to simulate vehicle movements, with RSUs randomly deployed at junctions of road segments. We utilized Simulation of Urban Mobility (SUMO) \cite{behrisch2011sumo} to generate random routes for $160\sim 200$ vehicles for dense regions and $30\sim 60$ vehicles for sparse regions. Among these vehicles, $10\% \sim 20\%$ of the vehicles were considered malicious. For each malicious vehicle, $1 \sim 10$ Sybil vehicles were randomly generated. Each vehicle trajectory was assigned a random start time. The traverse time between two RSUs was kept over 15 seconds \cite{baza2020detecting}. The simulation parameters are summarized in Table \ref{Table_2}. 

\begin{table}
    \centering
    \caption{Simulation Parameters}
    \begin{tabular}{|l|p{3.5cm}|}
        \hline
        \textbf{Parameter} & \textbf{Value} \\
        \hline
        Number of vehicles $N$ & $160\sim200$ (dense region),  $30\sim60$ (sparse region) \\
        \hline
        Vehicle speed & 80km/h $\sim$ 90km/h \cite{dutta2013time}\\
        \hline
        Transmission range of a vehicle & 250m \cite{dutta2013time}\\
        \hline
       Percentage of malicious vehicles $M$ & 10\% $\sim$ 20\% \\
      \hline
        Simulation time & 1000 sec \\
        \hline
    \end{tabular}
    \label{Table_2}
\end{table}

We compared the performance of the proposed scheme with that of Baza et al.'s scheme \cite{baza2020detecting}, denoted as PoW\_Scheme here after. The metrics used to assess the performance of both schemes are presented in Table \ref{table_3}. Experiments were conducted under different network conditions, including sparse and dense regions, with varying numbers of vehicles and different percentages of malicious vehicles.

We implemented both schemes using the Python programming language. For the proposed scheme, we recorded the latitude and longitude of vehicles at different timestamps. In contrast, for the PoW\_Scheme, we logged the interactions between vehicles and RSUs at different timestamps. For each experimental setup, the simulation was executed 10 times,  and we recorded the average results of all runs for analysis.

All experiments were conducted on a workstation running a 64-bit operating system on an x64-based architecture. The system was equipped with an 11th Generation Intel Core i7 processor operating at 2.30 GHz, 16 GB of RAM, and an NVIDIA GPU with 4 GB of dedicated memory. 

\begin{table}
\caption{Performance Metrics}
\scalebox{.77}{
\begin{tabular}{|p{1.7cm}|p{5cm}|p{3cm}|}
\hline
\textbf{Metric} & \textbf{Description} &  \\
\hline
False Positive Rate (FPR) & Proportion of legitimate vehicles that is identified as malicious or Sybil vehicles. 
\begin{center}
 $FPR = \frac{FP}{FP+TN} \times 100$ \end{center}\vspace{.02em} & \multirow{6}{3cm}{\vspace{1.5em} \\FP = incorrectly predicted as  malicious or Sybil vehicles \\
 FN = incorrectly detected as nonmalicious vehicles \\
 TN = correctly identified as nonmalicious vehicles \\
 TP = correctly predicted as malicious or Sybil vehicles} 
 \\
\cline{1-2}
False Negative Rate (FNR) & Percentage of malicious and Sybil vehicles that remain undetected. \begin{center}  
  $FNR = \frac{FN}{TP + FN} \times 100$\end{center}\vspace{0.02em}
   & \\
\cline{1-2}
Detection Rate &
   Percentage of actual malicious and Sybil vehicles that is correctly identified. \begin{center}
   $Detection Rate$ = $\frac{TP}{TP + FN} \times 100$ \end{center}\vspace{0.02em}
   & \\
\cline{1-2}
Accuracy & Percentage of correctly identified legitimate, and malicious and Sybil vehicles to the total number of vehicles. \begin{center}
   $Accuracy$ = $\frac{TP + TN}{TP + TN + FP + FN} \times 100$ \end{center}\vspace{0.02em}
   & \\
   \cline{1-2}
F1-Score & Combines TP, FP, and FN into a single metric\begin{center}
$F1$ = $\frac{2TP}{2TP + FP + FN}$
\end{center}
\vspace{0.02em} & \\
   \hline
    Latency & Time to detect a Sybil attack. & \\
\hline
\end{tabular}}
\label{table_3}
\end{table}

\begin{figure*}  
    \centering
    \subfloat[False Positive Rate]{
        \includegraphics[width=0.33\textwidth]{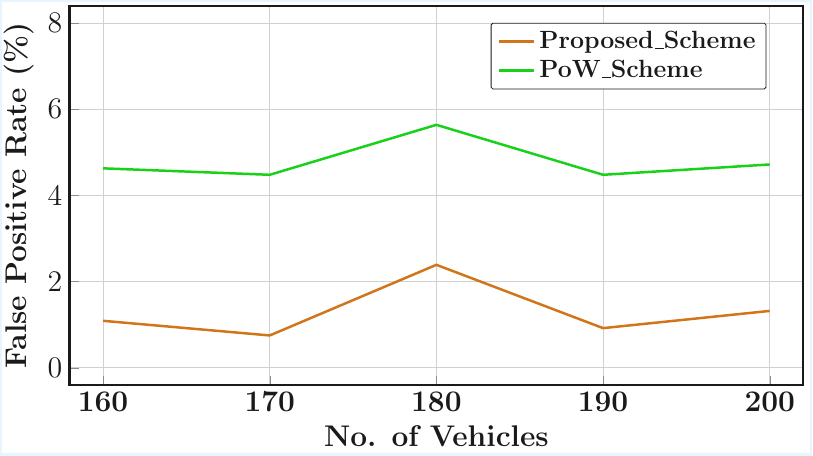}
        \label{6a}
    }
    \subfloat[False Negative Rate]{
        \includegraphics[width=0.33\textwidth]{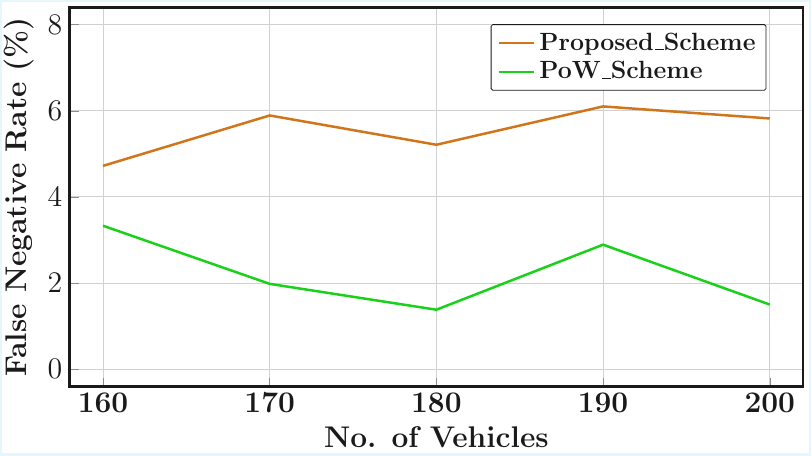}
        \label{6b}
    }
    \subfloat[Detection Rate]{
        \includegraphics[width=0.33\textwidth]{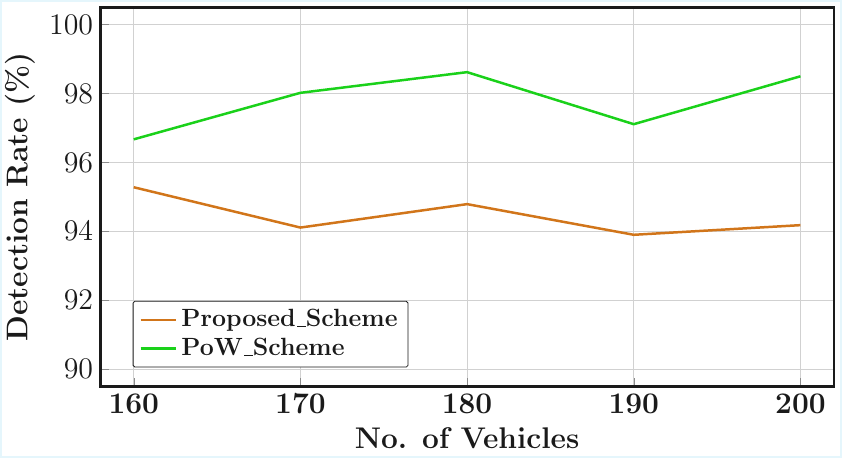}
        \label{6c}
    }
    \hfill
    \subfloat[Accuracy]{
        \includegraphics[width=0.33\textwidth]{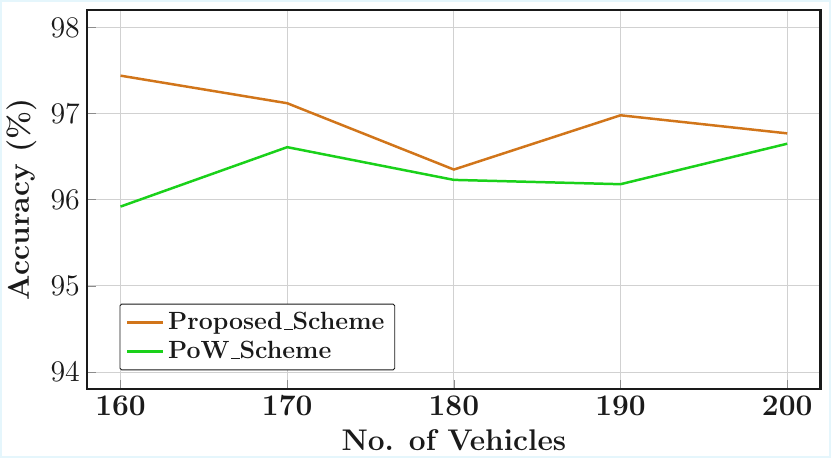}
        \label{6d}
    }
    \subfloat[F1 Score]{
        \includegraphics[width=0.33\textwidth]{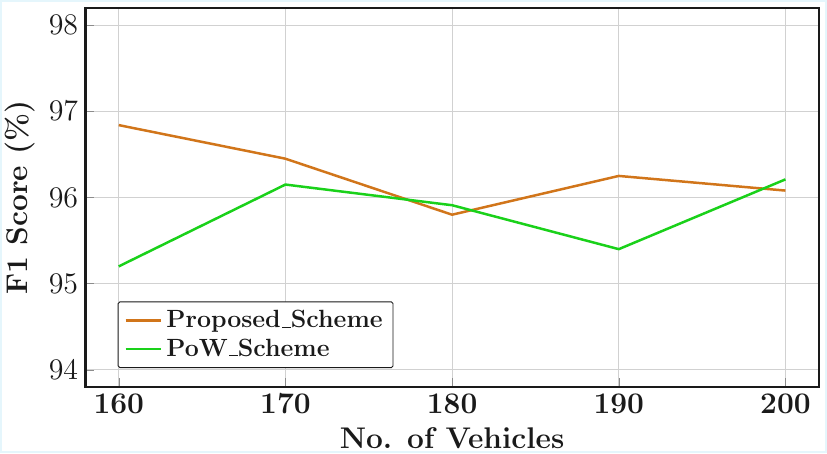}
        \label{6e}
    }
    \caption{Performance analysis for different values of $N$ in dense region when $M=10\%$.}
    \label{fig_6}
\end{figure*}

\subsection{Performance Evaluation in Dense Region}

\subsubsection{Impact of Network Size}\label{network_size_impact}

We evaluated the performance of both schemes with increasing number of vehicles in a dense network environment. Figure \ref{fig_6} demonstrates the performance where vehicles range in $160 \sim 200$, maintaining a fixed 10\% malicious vehicles. 

Figure \ref{6a} shows that the proposed scheme exhibits a better FPR in the range of $1\%$ to slightly more than $2\%$, compared to $4\%\sim5\%$ for the PoW\_scheme. The PoW\_Scheme uses check window size and trajectory length limit heuristics, which ensure that Sybil vehicles traveling close to each other fail the exclusion test and are detected as Sybil vehicles. However, this test also causes many legitimate vehicles to be detected as Sybil vehicles since it is possible for normal vehicles to be present at the same RSU within the time size window and have a combined RSU length less than the trajectory length limit. In contrast, the proposed scheme forms clusters based on the similarity of the entire trajectory of the vehicles. Sybil vehicles tend to follow similar routes, which leads their trajectories to cluster together. On the contrary, the legitimate vehicles' trajectories are generally distinct and do not overlap with other trajectories, decreasing the possibility of forming clusters. As shown in Fig. \ref{6a}, the proposed scheme achieves an improvement of approximately 72\%  
in reducing FPR over the PoW\_Scheme when $N$=200. 

As shown in Fig. \ref{6b}, the proposed scheme exhibits FNR in the range of $4\%\sim6\%$, whereas the PoW\_Scheme results in FNR in the range $1\%\sim4\%$. The PoW\_Scheme uses check window size and trajectory length limit heuristics, which ensure that Sybil vehicles fail the exclusion test and are detected as Sybil vehicles. In the proposed scheme, although Sybil vehicles generally follow similar trajectories, some of them show more spatial dispersion than others. While a subset of Sybil vehicles forms compact and dense clusters, a few are located slightly farther apart. Since the $\varepsilon$ parameter is determined dynamically using the elbow method, it is primarily optimized to capture the densest clusters in the similarity space. Consequently, the more dispersed Sybil vehicles may fall outside the selected neighborhood radius and are not grouped into the Sybil cluster. This leads to certain Sybil vehicles being misclassified as legitimate vehicles. Figure \ref{6b} shows that the PoW\_Scheme reduces the FNR by nearly 74\%  
compared to the proposed scheme when $N$=200.

Due to the higher FNR, the proposed scheme shows a lower detection rate as shown in Fig. \ref{6c}. The detection rate for the proposed scheme ranges between $94\%\sim95\%$,  whereas for the PoW\_Scheme it ranges in $97\%$ to slightly more that $98\%$. The PoW\_Scheme attains nearly 5\% improvement in detection rate compared to the proposed scheme when $N$=200 as shown in Fig. \ref{6c}.

Figure \ref{6d} shows that both the proposed and the PoW\_Schemes achieve an average accuracy of  $97\%$ and $96\%$, respectively. Both schemes achieve nearly the same performance due to having almost similar FPR and FNR. 

For the same reasons, the F1 score for the proposed scheme reaches almost 97\%, while PoW\_Scheme's score peaks at just over 96\% as shown in Fig. \ref{6e}.

Although the PoW\_Scheme demonstrates better performance for FNR and detection rate, it is important to note that these results were obtained using optimally tuned heuristic parameters for each specific setup. In practice, this scheme needs manual tuning based on ground-truth data for every new scenario to achieve optimal performance. In contrast, the proposed scheme does not depend on manually calibrated parameters. Instead, it employs an unsupervised approach to detect Sybil vehicles, eliminating the need for ground-truth data during parameter selection. As a result, the proposed method can adapt to different environments without requiring environment-specific tuning, making it more practical and scalable for real-world deployment.
\begin{figure*}  
    \centering
    \subfloat[False Positive Rate]{
        \includegraphics[width=0.33\textwidth]{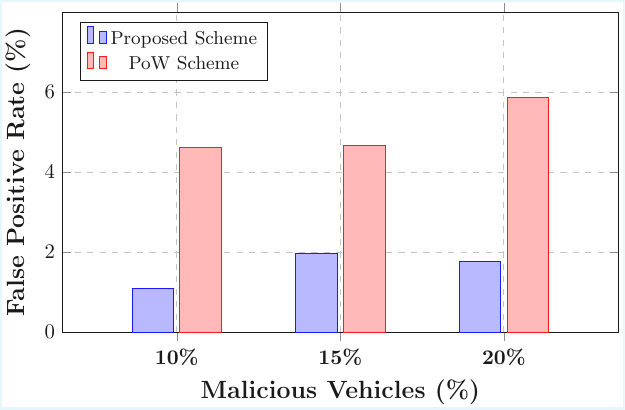}
        \label{7a}
    }
    \subfloat[False Negative Rate]{
        \includegraphics[width=0.33\textwidth]{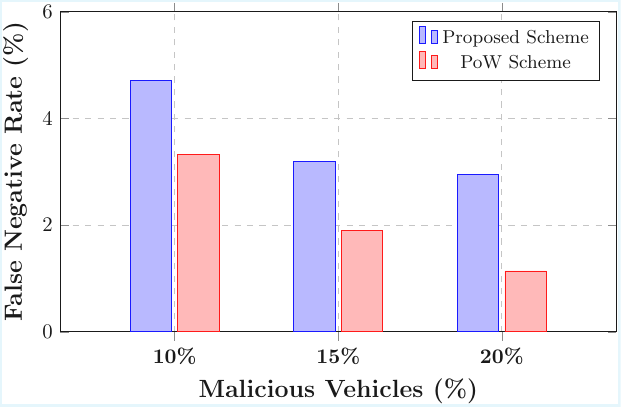}
        \label{7b}
    }
    \subfloat[Detection Rate]{
        \includegraphics[width=0.33\textwidth]{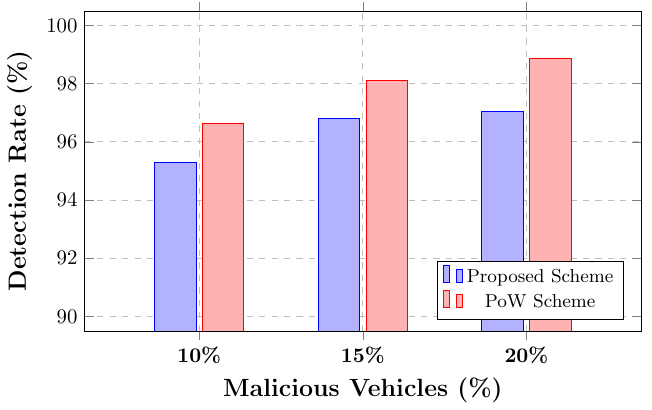}
        \label{7c}
    }
    \hfill
    \subfloat[Accuracy]{
        \includegraphics[width=0.33\textwidth]{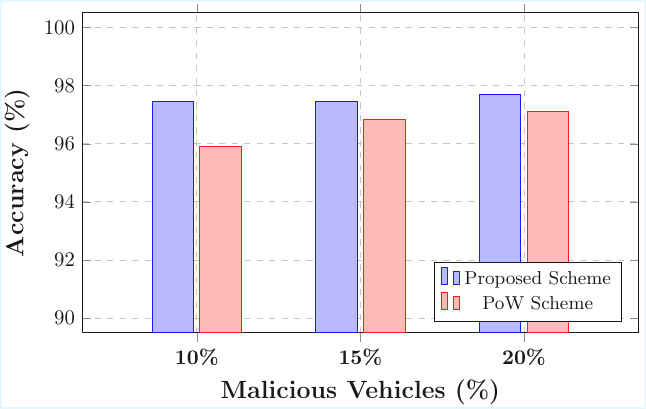}
        \label{7d}
    }
    \subfloat[F1 Score]{
        \includegraphics[width=0.33\textwidth]{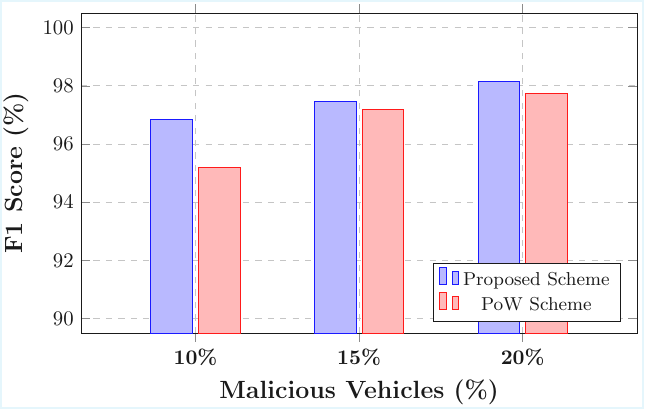}
        \label{7e}
    }
    \caption{Performance analysis for different percentages of $M$ in dense region when $N=160$.}
    \label{fig_7}
\end{figure*}

\subsubsection{Impact of Malicious Vehicles} Figure \ref{fig_7} demonstrates the impact of increasing percentage of malicious vehicles $M$ on both schemes while keeping the number of vehicles $N$ fixed to 160. 

Figure \ref{7a} shows that the FPR increases almost linearly for the PoW\_Scheme, reaching almost 6\% when $M$=20\%. In contrast, the FPR for the proposed scheme remains below 2\% as $M$ increases. With more Sybil vehicles present, the number of Sybil–normal vehicle trajectory pairs also increases in the PoW\_Scheme. Many of these pairs fail the exclusion test because they appear within the same RSU during the check-size window and produce a combined trajectory length below the predefined threshold. Hence, more normal vehicles are misclassified as Sybil vehicles, leading to a higher FPR. For the proposed scheme, the increase in malicious vehicles does not significantly affect the FPR. This is because, the presence of additional Sybil vehicles strengthens the density of existing Sybil clusters. Consequently, the dynamically calculated $\varepsilon$ either remains stable or decreases slightly, reducing the likelihood of normal vehicles being incorrectly included in Sybil clusters. As shown in Fig. \ref{7a}, the proposed scheme reduces the FPR by 70\% compared to the PoW\_Scheme at $M$=20\%.

Figure \ref{7b} shows that the FNR decreases with the higher values of $M$ for both the proposed scheme and PoW\_Scheme, while the PoW\_Scheme exhibits less FNR compared to the proposed scheme. As shown in Fig. \ref{7b}, the PoW\_Scheme reduces the FNR by nearly 61\% compared to the proposed scheme at $M$=20\%. As the percentage of Sybil vehicles increases in the network, more of them fall within the neighborhood radius $\varepsilon$ and form dense clusters in the proposed scheme. Consequently, a larger proportion of Sybil vehicles is correctly grouped, leading to a reduced FNR with increasing percentage of $M$. Nevertheless, the proposed scheme has a higher FNR than the PoW\_Scheme for the reasons discussed in Section \ref{network_size_impact}. 

As shown in Fig. \ref{7c}, the detection rate escalates with the increasing percentage of $M$ in both schemes. This finding is aligned with the trends in FNR shown in Fig. \ref{7b}, where FNR decreases with the increasing percentage of $M$. The detection rates for the PoW\_Scheme and the proposed scheme approach nearly 99\% and 97\%, respectively, at $M$ = 20\%. As shown in Fig. \ref{7c}, the PoW\_Scheme achieves an improvement of nearly 2\% over the proposed scheme at $M$ = 20\%.  

Figure \ref{7d} shows that the proposed scheme attains almost similar accuracy for all values of $M$ (nearly 98\%), whereas accuracy for the PoW\_Scheme improves with increasing $M$, reaching approximately 97\%. 

The F1 score increases linearly for both schemes with higher values of $M$ as shown in Fig. \ref{7e}. The proposed scheme and PoW\_Scheme achieve an F1 score of almost 98\% when $M$ = 20\%, where PoW\_Scheme slightly deviates from the proposed scheme.

\subsection{Performance Evaluation in Sparse Region}

\subsubsection{Impact of Network Size}\label{network_size_sparse}
\begin{figure*}  
    \centering
    \subfloat[False Positive Rate]{
        \includegraphics[width=0.33\textwidth]{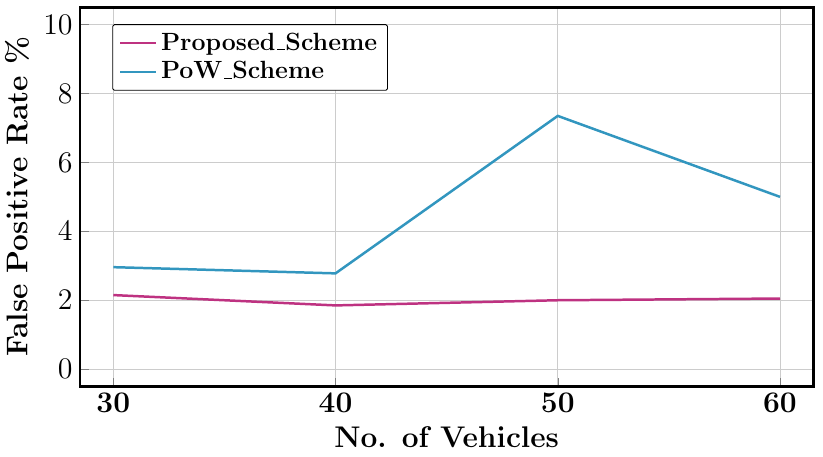}
        \label{8a}
    }
    \subfloat[False Negative Rate]{
        \includegraphics[width=0.33\textwidth]{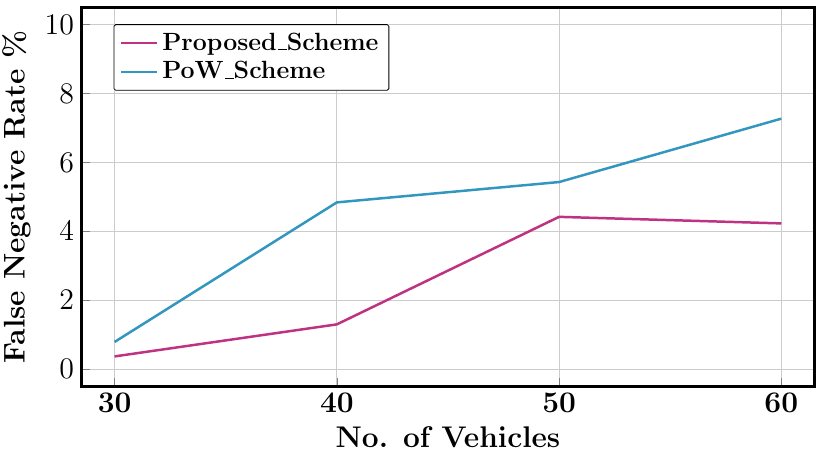}
        \label{8b}
    }
    \subfloat[Detection Rate]{
        \includegraphics[width=0.33\textwidth]{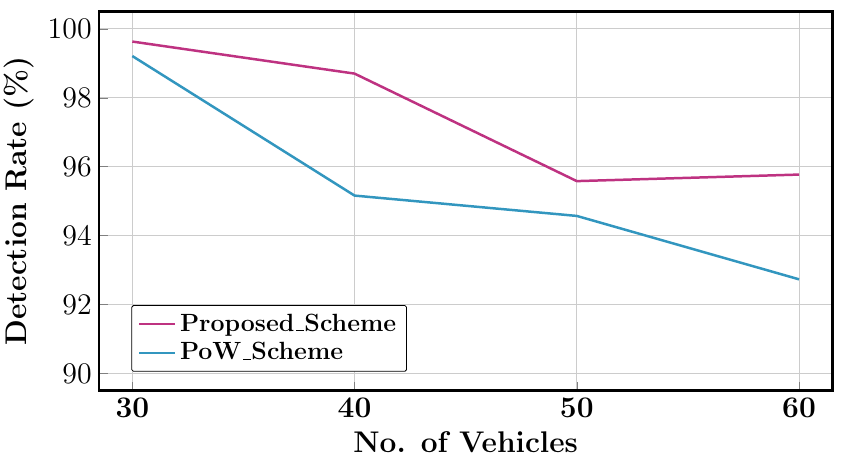}
        \label{8c}
    }
    \hfill
    \subfloat[Accuracy]{
        \includegraphics[width=0.33\textwidth]{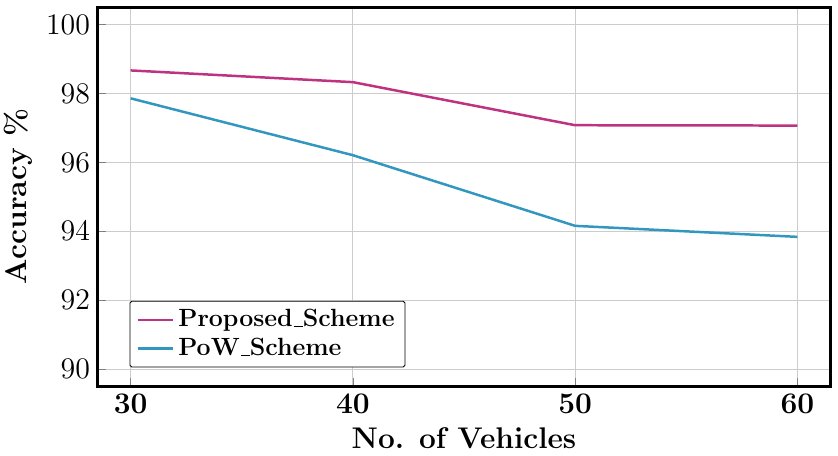}
        \label{8d}
    }
    \subfloat[F1 Score]{
        \includegraphics[width=0.33\textwidth]{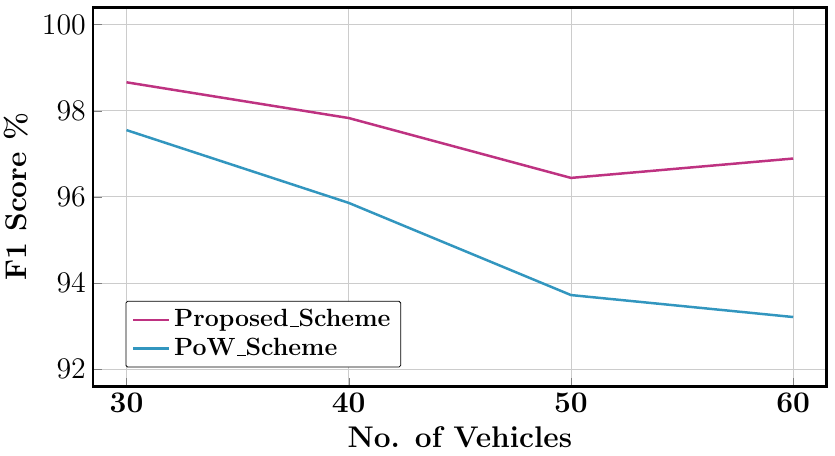}
        \label{8e}
    }
    \caption{Performance analysis for different values of $N$ in sparse environment when $M=10\%$.}
    \label{fig_8}
\end{figure*}

Unlike the dense region, the proposed scheme demonstrates superior results across all metrics in comparison to the PoW\_Scheme in a sparse setting, as shown in Fig. \ref{fig_8}. Due to fewer vehicles in a sparse region (30 $\sim$ 60 vehicles), the possibility that a legitimate vehicle's trajectory aligns with the trajectory of a Sybil vehicle decreases. As a result,  the proposed scheme achieves better results in  FPR (2\%), and FNR (slightly more than 4\%), leading to higher detection rate (approximately 96\%), accuracy (approximately 97\%), and F1 score (almost 97\%) at $N$=60 as shown in Figs. \ref{8a}, \ref{8b}, \ref{8c}, \ref{8d}, and \ref{8e}, respectively. Conversely, the PoW\_Scheme is sensitive to the heuristic values, particularly to the check size window. Some pairs of Sybil vehicles may fall outside the range of the check size window, falsely passing the exclusion test, and being classified as legitimate. This increases the PoW\_Scheme's FNR and reduces the detection rate. Additionally, the PoW\_Scheme achieves a higher FPR for the reasons discussed in Section \ref{network_size_impact}. The proposed scheme obtains an improvement of 59\%, 42\%, 3\%, 3\%, and 4\% for FPR, FNR, detection rate, accuracy, and F1 score, respectively, over the PoW\_Scheme when $N$=60.
\begin{figure*}  
    \centering
    \subfloat[False Positive Rate]{
        \includegraphics[width=0.33\textwidth]{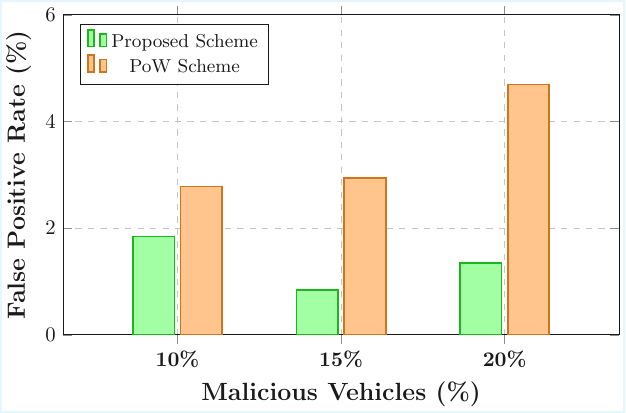}
        \label{9a}
    }
    \subfloat[False Negative Rate]{
        \includegraphics[width=0.33\textwidth]{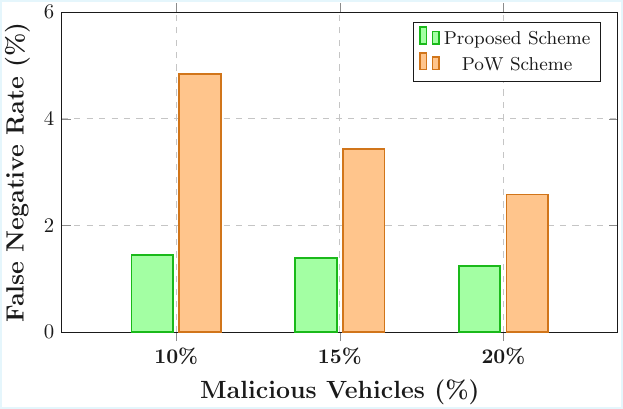}
        \label{9b}
    }
    \subfloat[Detection Rate]{
        \includegraphics[width=0.33\textwidth]{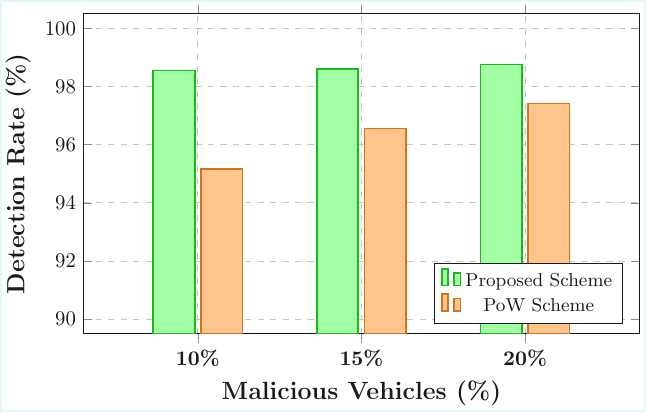}
        \label{9c}
    }
    \hfill
    \subfloat[Accuracy]{
        \includegraphics[width=0.33\textwidth]{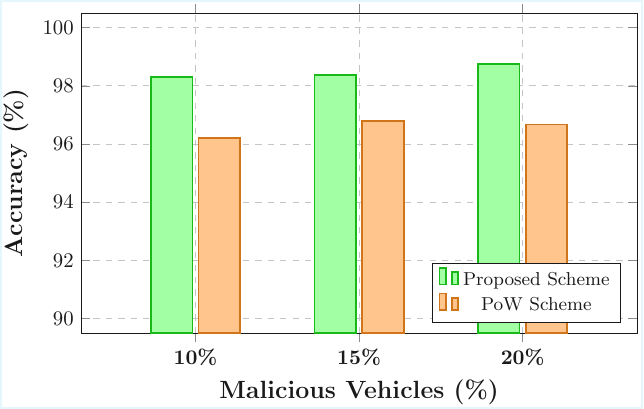}
        \label{9d}
    }
    \subfloat[F1 Score]{
        \includegraphics[width=0.33\textwidth]{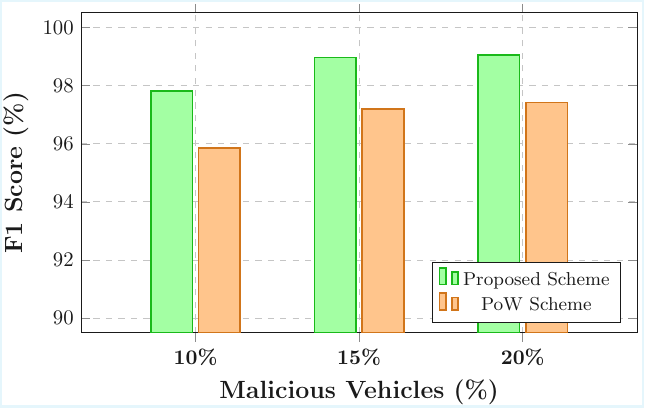}
        \label{9e}
    }
    \caption{Performance analysis for different percentages of $M$ in sparse region when $N=40$.}
    \label{fig_9}
\end{figure*}

\subsubsection{Impact of Malicious Vehicles}
Figure \ref{fig_9} shows the performance of both schemes for increasing percentages of $M$ when $N$=40. The performance trends shown in Fig. \ref{fig_9} are completely aligned with the reasons discussed in Section \ref{network_size_sparse}. The proposed scheme achieves an improvement of nearly 72\% for FPR and 52\% for FNR at $M$=20\% as shown in Fig. \ref{9a}, and Fig. \ref{9b}, respectively. It obtains a detection rate, accuracy, and F1 score of approximately 99\% at $M$=20\%, as shown in Figs. \ref{9c}, \ref{9d}, and \ref{9e}.

\begin{figure*}  
    \centering
    \subfloat[Dense region]{
        \includegraphics[width=0.45\textwidth]{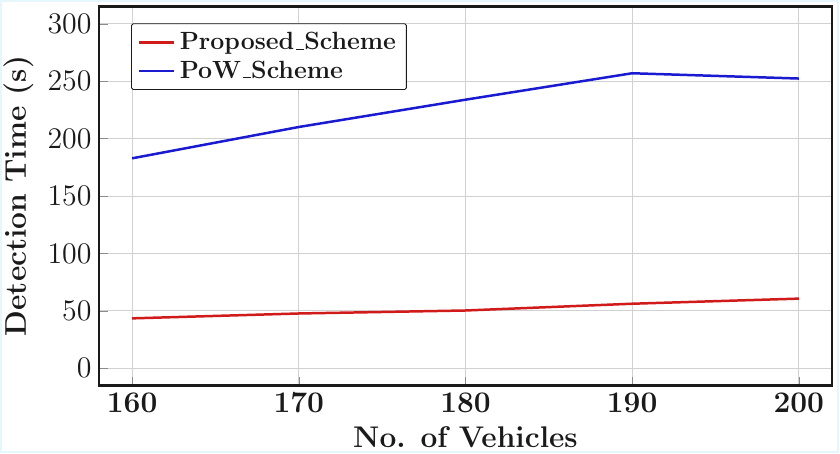}
        \label{10a}
    }
    \subfloat[Sparse region]{
        \includegraphics[width=0.45\textwidth]{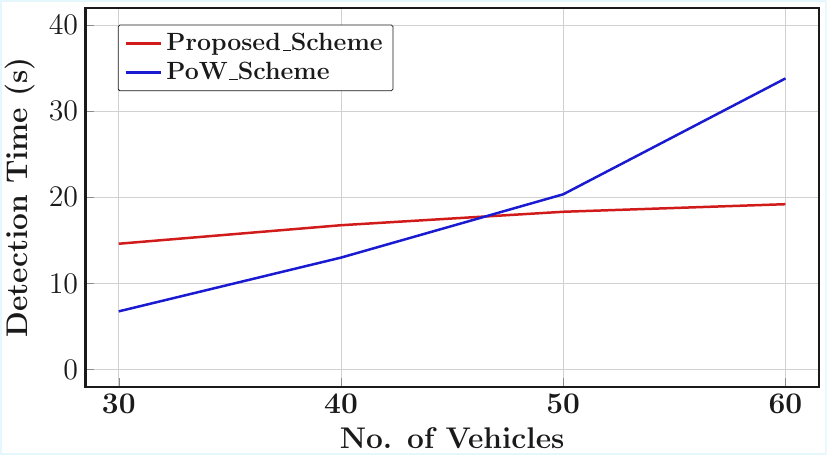}
        \label{10b}
    }
    
    \caption{Detection time for different values of $N$ when $M$=10\%.}
    \label{fig_10}
\end{figure*}

\subsection{Detection time}
The detection time for the PoW\_Scheme includes the time to execute both heuristics, compute the graph, and find the clique. In contrast, the detection time in the proposed scheme comprises the time to calculate the similarity matrix using the discrete Fr\'echet distance, determine the optimal $\varepsilon$, and cluster the trajectories.

Figure \ref{fig_10} shows the detection time for the proposed scheme in comparison to the PoW\_Scheme for different values of $N$ when $M$=10\%. The proposed scheme shows better results in the dense region. As shown in Fig. \ref{10a}, it reduces the detection time by approximately 76\% compared to the PoW\_Scheme when $N=200$. In contrast, the PoW\_Scheme shows better detection time for low values of $N$ for sparse region as shown in Fig. \ref{10b}. However, for higher values of $N$ ($N>$46), the proposed scheme outperforms the PoW\_Scheme shown in Fig. \ref{10b}. Additionally, the proposed scheme maintains a detection time less than 20 seconds for all values of $N$ in a sparse region. It achieves an improvement of almost 43\% compared to the PoW\_Scheme when $N$ = 60.

In the proposed scheme, the time complexity for computing the discrete Fr\'echet distance between trajectories with $p$ and $q$ points is $\mathcal{O}(pq)$ \cite{EiterM94}. The time complexity to compute the similarity matrix consisting of all pairs of trajectories is $\mathcal{O}(N^2pq)$, where $N$ is the number of vehicles. Computing the optimal $\varepsilon$ value requires sorting the average $k-1$ nearest neighbor distance of vehicles. Sorting each row of the similarity matrix takes $\mathcal{O}(N log N)$. Hence, the total time complexity for computing the optimal $\varepsilon$ takes $\mathcal{O}(N^2log N)$. Finally,  executing the DBSCAN algorithm on the precomputed similarity matrix needs $\mathcal{O}(N^2)$ time \cite{gunawan2013faster}. The overall time complexity is predominated by the formation of the similarity matrix using discrete Fr\'echet distance, taking $\mathcal{O}(N^2pq)$ time.

Note that in the PoW\_Scheme, the time required to determine the optimal heuristic values - check window size and trajectory length limit - was not included in the detection process during the experiment. This process involves varying one heuristic while keeping the other constant, significantly increasing the detection time. 

\section{Conclusions}\label{conclusion}
We have proposed a novel approach for detecting Sybil attacks in VANETs, where a vehicle’s trajectory is constructed from continuous GPS readings, with RSUs acting as trusted entities for verification. These verified trajectories are fed into the DBSCAN clustering algorithm to identify potential Sybil vehicles. Our proposed scheme significantly reduces the FPR and eliminates the need for manual parameter tuning using ground-truth data. By removing computationally intensive puzzle-solving, this method prevents malicious vehicles from gaining an unfair advantage due to differences in computational resources. Additionally, our approach is effective in both dense and sparse regions. Experimental results show that the proposed scheme reduces the FPR by almost 72\% in both dense and sparse regions. In sparse regions, it also lowers the FNR by 52\% and exceeds existing works by obtaining a detection rate of 99\%. Additionally, it decreases the detection time by almost 76\% in dense regions. In the future, we plan to extend this work to scenarios involving untrusted RSUs. Furthermore, incorporating a more advanced location verification mechanism and developing a stronger mechanism to capture the complex and dynamic patterns of vehicle movement are interesting directions of research work.

\bibliographystyle{ieeetr}
\bibliography{main}

@article{yu2013detecting,
  title={Detecting {Sybil} attacks in {VANETs}},
  author={Yu, Bo and Xu, Cheng-Zhong and Xiao, Bin},
  journal={J. Parallel Distrib. Comput.},
  volume={73},
  number={6},
  pages={746--756},
  year={2013},
  publisher={Elsevier}
}

@article{qu2015security,
  title={A security and privacy review of {VANETs}},
  author={Qu, Fengzhong and Wu, Zhihui and Wang, Fei-Yue and Cho, Woong},
  journal={IEEE Trans. Intell. Transp. Syst.},
  volume={16},
  number={6},
  pages={2985--2996},
  year={2015},
  publisher={IEEE}
}

@article{zhang2023detection,
  title={Detection method to eliminate {Sybil} attacks in {Vehicular Ad-hoc Networks}},
  author={Zhang, Zhaoyi and Lai, Yingxu and Chen, Ye and Wei, Jingwen and Wang, Yuhang},
  journal={Ad Hoc Networks},
  volume={141},
  pages={103092},
  year={2023},
  publisher={Elsevier}
}

@article{hasan2020securing,
  title={Securing vehicle-to-everything {(V2X)} communication platforms},
  author={Hasan, Monowar and Mohan, Sibin and Shimizu, Takayuki and Lu, Hongsheng},
  journal={IEEE Trans. Intell. Vehicles},
  volume={5},
  number={4},
  pages={693--713},
  year={2020},
  publisher={IEEE}
}

@article{sakiz2017survey,
  title={A survey of attacks and detection mechanisms on intelligent transportation systems: {VANETs} and {IoV}},
  author={Sakiz, Fatih and Sen, Sevil},
  journal={Ad Hoc Networks},
  volume={61},
  pages={33--50},
  year={2017},
  publisher={Elsevier}
}

@inproceedings{zhang2015exploiting,
  title={Exploiting mobile social behaviors for {Sybil} detection},
  author={Zhang, Kuan and Liang, Xiaohui and Lu, Rongxing and Yang, Kan and Shen, Xuemin Sherman},
  booktitle={Proc. IEEE Conf. Comput. Commun. (INFOCOM)},
  pages={271--279},
  year={2015},
  organization={IEEE}
}

@inproceedings{rabieh2015cross,
  title={Cross-layer scheme for detecting large-scale colluding {Sybil} attack in {VANETs}},
  author={Rabieh, Khaled and M. E. A Mahmoud, Mohamed and Guo, Terry N and Younis, Mohamed},
  booktitle={Proc. IEEE Int. Conf. Commun. (ICC)},
  pages={7298--7303},
  year={2015},
  organization={IEEE}
}

@inproceedings{pattanayak2020novel,
  title={A novel approach to detection of and protection from {Sybil} attack in {VANET}},
  author={Pattanayak, Binod Kumar and Pattnaik, Omkar and Pani, Sasmita},
  booktitle={Advances in Intell. Comput. Commun.},
  pages={240--247},
  year={2020},
  organization={Springer}
}

@inproceedings{subramanian2020decentralized,
  title={Decentralized device authentication model using the trust score and blockchain technology for dynamic networks},
  author={Subramanian, Venkatesan and Rajendra, Yuvaraj and Sahai, Shubham and Shukla, Sandeep K},
  booktitle={Proc. Int. Conf. Blockchain (Blockchain)},
  pages={116--125},
  year={2020},
  organization={IEEE}
}

@article{chang2011footprint,
  title={Footprint: detecting {Sybil} attacks in urban vehicular networks},
  author={Chang, Shan and Qi, Yong and Zhu, Hongzi and Zhao, Jizhong and Shen, Xuemin},
  journal={IEEE Trans. Parallel Distrib. Syst.},
  volume={23},
  number={6},
  pages={1103--1114},
  year={2012},
  publisher={IEEE}
}

@article{baza2020detecting,
  title={Detecting {Sybil} attacks using {Proofs} of {Work} and location in {VANETs}},
  author={Baza, Mohamed and Nabil, Mahmoud and Mahmoud, Mohamed M. E. A and Bewermeier, Niclas and Fidan, Kemal and Alasmary, Waleed and Abdallah, Mohamed},
  journal={IEEE Trans. Dependable Secure Comput.},
  volume={19},
  number={1},
  pages={39--53},
  year={2022},
  publisher={IEEE}
}

@article{rajendra2024sybil,
  title={{Sybil} attack detection in ultra-dense {VANETs} using verifiable delay functions},
  author={Rajendra, Yuvaraj and Subramanian, Venkatesan and Shukla, Sandeep Kumar},
  journal={Peer-to-Peer Netw. Appl.},
  volume={17},
  pages={1645–1666},
  year={2024},
  publisher={Springer}
}

@inproceedings{dutta2013time,
  title={A time-series clustering approach for {Sybil} attack detection in {Vehicular Ad hoc Networks}},
  author={Dutta, Neelanjana and Chellappan, Sriram},
  booktitle={Proc. Int. Conf. Advances Veh. Syst. Technol. Appl},
  pages={35--40},
  year={2013}
}

@article{li2021trajectory,
  title={Trajectory as an {Identity}: Privacy-Preserving and {Sybil}-Resistant Authentication for {Internet of Vehicles}},
  author={Li, Jiangtao and Song, Zhaoheng and Li, Yufeng and Cao, Chenhong and He, Yuanhang},
  journal={Secur. Commun. Netw.},
  volume={2021},
  number={1},
  pages={8251697},
  year={2021},
  publisher={Wiley Online Library}
}

@inproceedings{deng2020dbscan,
  title={{DBSCAN} clustering algorithm based on density},
  author={Deng, Dingsheng},
  booktitle={Proc. 7th Int. Forum Elect. Eng. Automat. (IFEEA)},
  pages={949--953},
  year={2020},
  organization={IEEE}
}

@phdthesis{cavallaro2021analysis,
    title={Analysis of large {GPS} trajectories datasets via multi-agent techniques},
    author={Cavallaro, Claudia},
    school ={Universit{\`a} Degli Studi Di Catania} ,
    year = {2021}
}

@article{su2020survey,
  title={A survey of trajectory distance measures and performance evaluation},
  author={Su, Han and Liu, Shuncheng and Zheng, Bolong and Zhou, Xiaofang and Zheng, Kai},
  journal={The {VLDB} Journal},
  volume={29},
  number={1},
  pages={3--32},
  year={2020},
  publisher={Springer}
}

@techreport{EiterM94,
  author    = {Thomas Eiter and Heikki Mannila},
  title     = {Computing discrete {Fr{\'e}chet} distance},
  institution = {Technical University of Vienna},
  year      = {1994},
  number    = {CD-TR 94/64}
}

@inproceedings{ester1996density,
  title={A density-based algorithm for discovering clusters in large spatial databases with noise},
  author={Ester, Martin and Kriegel, Hans-Peter and Sander, J{\"o}rg and Xu, Xiaowei},
  booktitle={Proc. Int. Conf. Knowl. Discovery Data Mining},
  pages={226--231},
  year={1996}
}

@article{kara2020effect,
  title={Effect of {RSU} placement on autonomous vehicle {V2I} scenarios},
  author={Kara, Bar{\i}{\c{s}} and {\"O}zg{\"o}vde, Bahri Atay},
  journal={Balkan J. Elect. Comput. Eng.},
  volume={8},
  number={3},
  pages={272--284},
  year={2020},
  publisher={MUSA YILMAZ}
}

@inproceedings{wei2019rsu,
  title={{RSU} beacon aided trust management system for location privacy-enhanced {VANETs}},
  author={Wei, Yu-Chih and Chen, Yi-Ming and Wu, Wei-Chen and Chu, Ya-Chi},
  booktitle={Proc. Frontier Comput.: Theory, Technologies and Appl. (FC 2018)},
  pages={1913--1924},
  year={2019},
  organization={Springer}
}

@article{kperiyarselvam2024,
    title={Advancing Road Safety through Cloud Based {RSU} Solutions for Smart {Internet of Vehicles}},
    author={ K. Periyarselvam and G. Akash and V Harish and V. Thunaivan },
    journal={J. ISMAC: The J. IoT in Social, Mobile, Anal., Cloud},
    volume={6},
    number={2},
    year={ 2024 } 
}

@article{mershad2012roamer,
  title={{ROAMER}: Roadside Units as message routers in {VANETs}},
  author={Mershad, Khaleel and Artail, Hassan and Gerla, Mario},
  journal={Ad Hoc Networks},
  volume={10},
  number={3},
  pages={479--496},
  year={2012},
  publisher={Elsevier}
}

@article{kaushik2022transmit,
  title={Transmit range adjustment using {Artificial Intelligence} for enhancement of location privacy and data security in service location protocol of {VANET}},
  author={Kaushik, Shivkant and Poonia, Ramesh Chandra and Khatri, Sunil Kumar and Samanta, Debabrata and Chakraborty, Partha},
  journal={Wireless Commun. and Mobile Comput.},
  volume={2022},
  number={1},
  pages={9642774},
  year={2022},
  publisher={Wiley Online Library}
}

@article{zhao2016high,
  title={High-precision vehicle navigation in urban environments using an {MEM's} {IMU} and single-frequency {GPS} receiver},
  author={Zhao, Sheng and Chen, Yiming and Farrell, Jay A},
  journal={IEEE Trans. Intell. Transp. Syst.},
  volume={17},
  number={10},
  pages={2854--2867},
  year={2016},
  publisher={IEEE}
}

@article{liu2024joint,
  title={Joint {RSU} and agent vehicle cooperative localization using {mmWave} sensing},
  author={Liu, Yuanxin and Li, Demin and Chen, Xuemin},
  journal={Physical Commun.},
  volume={67},
  pages={102535},
  year={2024},
  publisher={Elsevier}
}

@inproceedings{gutschlag2022generalized,
  title={On the generalized {Fr{\'e}chet distance} and its applications},
  author={Gutschlag, Theodor and Storandt, Sabine},
  booktitle={Proc. 30th Int. Conf. on Advances in Geographic Inf. Syst.},
  pages={1--10},
  year={2022}
}

@article{schubert2017dbscan,
  title={{DBSCAN} revisited, revisited: why and how you should (still) use {DBSCAN}},
  author={Schubert, Erich and Sander, J{\"o}rg and Ester, Martin and Kriegel, Hans Peter and Xu, Xiaowei},
  journal={ACM Trans. on Database Syst. (TODS)},
  volume={42},
  number={3},
  pages={1--21},
  year={2017},
  publisher={Acm New York, NY, USA}
}

@inproceedings{rahmah2016determination,
  title={Determination of optimal epsilon (eps) value on {DBSCAN} algorithm to clustering data on peatland hotspots in sumatra},
  author={Rahmah, Nadia and Sitanggang, Imas Sukaesih},
  booktitle={IOP Conf. series: earth and environmental Sci.},
  volume={31},
  pages={012012},
  year={2016},
  organization={IoP Publishing}
}

@misc{map2017open,
  title={Open street map},
  howpublished = {\url{https://openstreetmap.org}},
  note={Accessed on 2025-04-15}
}

@inproceedings{behrisch2011sumo,
  title={{SUMO}--simulation of urban mobility: an overview},
  author={Behrisch, Michael and Bieker, Laura and Erdmann, Jakob and Krajzewicz, Daniel},
  booktitle={Proc. 3rd Int. Conf. Adv. Syst. Simul.},
  year={2011}
}

@mastersthesis{gunawan2013faster,
    title={A faster algorithm for {DBSCAN}},
    author={Gunawan, Ade},
    school ={Eindhoven University of Technology} ,
    year = {2013}
}
\end{document}